\documentclass[twocolumn,epsfig,showpacs]{revtex4}
\usepackage{epsfig,color}

\newcommand{\kp}{\mbox{K$^+$~}}
\newcommand{\km}{\mbox{K$^-$~}}

\newcommand{\AGeV}[1][ ]{$A$~GeV{#1}}

\begin{document}
\title{How to determine experimentally the \kp nucleus potential and the \kp N rescattering cross section in
a hadronic environment?}

\author{Ch. Hartnack$^1$ A. Sood$^1$ H. Oeschler$^2$ and
J. Aichelin$^1$ }
\address{
$^1$SUBATECH,
Laboratoire de Physique Subatomique et des
Technologies Associ\'ees \\University of Nantes - IN2P3/CNRS - Ecole des Mines
de Nantes \\
4 rue Alfred Kastler, F-44072 Nantes, Cedex 03, France\\
$^2$Institut f\"ur Kernphysik, Darmstadt University of Technology,
D-64298 Darmstadt, Germany\\}
\begin{abstract}
Comparing \kp spectra at low transverse momenta for different
symmetric collision systems at beam energies around 1 \AGeV allows
for a direct determination of both the strength of the \kp nucleus
potential as well as of the $K^+N$ rescattering cross section in a
hadronic environment. Other little known or unknown quantities
which enter the \kp dynamics, like the production cross sections
of \kp mesons or the hadronic equation of state, do not spoil this
signal as they cancel by using ratios of spectra. This procedure
is based on transport model calculations using the Isospin Quantum
Molecular Dynamics (IQMD) model which describes the available data
quantitatively.
\end{abstract}
\date{\today}

\maketitle

The change of the properties of mesons in dense hadronic matter
has theoretically been investigated since many years
\cite{Eichstaedt:2007zp} and an experimental verification is still
missing. The $t\rho$ approximation allows predicting the optical
potential of the mesons in low-densities matter by experimentally
measured phase shifts. At higher densities more sophisticated
approaches have to be employed and in the last two decades many
efforts have been made to investigate the properties of $\rho$,
$\omega$, \kp and \km mesons in matter
\cite{Eichstaedt:2007zp,Riek:2010gz,Korpa:2004ae,Tolos:2006ny}.
These calculations are complex  because most of the mesons can
form baryonic resonances which have other decay branches.
Therefore, coupled-channels calculations have to be employed and
the challenge has been met to calculate them self-consistently.
Nevertheless, the theoretical predictions launched by different
groups differ substantially because several of the quantities
which enter such calculations, like in-medium coupling constants
and the in-medium dressing of the different particles are only
poorly known. These uncertainties render the theoretical
prediction rather vague and it is highly desirably to have
experimental information of properties of mesons around and above
normal nuclear matter densities.

It is the purpose of this Letter to show that one
observable can be identified which allow for the determination of the
desired quantities by experiment for one of the mesons, the
\mbox{K$^+$}. This meson is for several reasons the ideal
candidate to start with. First of all, all nuclear matter
calculations show that \kp behaves in medium as a quasi particle
and can therefore be propagated like particles in simulation
programs. Second, the \mbox{K$^+$}, having a valence ${\bar s}$
quark, does not form baryonic resonances, and hence, its
interaction with the medium is relatively simple.

The change of the energy of a \kp,  $\omega({\bf{k}},\rho)$,  with momentum $\bf{k}$ and at a density
$\rho$ ~\cite{SchaffnerBielich:2000jy,Schaffner:1996kv},
\begin{equation}
\omega({\bf{k}},\rho)= \sqrt{\left( {\bf{k}-\Sigma_v} \right)^2+m^2 + m
\Sigma_s} \pm \Sigma_v^0 \label{schaf}
\end{equation}
 depends on the a scalar
self energy $\Sigma_s$ and a vector self energy ($\Sigma_v^0, {\bf
\Sigma_v})$. The scalar potential $\Sigma_s$ is related to the
$\sigma$ field which itself is related in a non-linear way to the
scalar density $\rho_s$. The vector potential is related to the
baryon density $\rho_B$. For details we refer to
Ref.~\cite{Schaffner:1996kv}. Knowing $\omega({\bf{k}},\rho)$, one
can describe the ``mass'' of the \kp and its variation in a
nuclear environment $\omega({\bf{k}}=0,\rho)$. Korpa and Lutz
\cite{Korpa:2004ae} have calculated this quantity using a
self-consistent Bethe-Salpeter equation. The result of this
approach is very similar of that of the mean field approach
~\cite{SchaffnerBielich:2000jy,Schaffner:1996kv} and yields the
linear relation
\begin{equation} \omega({\bf{k}}=0,\rho)= m_{\rm
K^+}(\rho) = m_{\rm K^+}(\rho=0) \, (1+ \alpha_{\rm K^+}
\frac{\rho}{\rho_0}) \label{kpmass}
\end{equation}
with $\alpha_{\rm K^+} = 0.08$.

In a nuclear environment, the properties of the observed \kp is
not only determined by the \kp nucleus potential but also by \kp N
collisions. In free space the cross section for this interaction,
$\sigma_{\rm{K^+p}}$, is about 12 mb, and similar for
$\sigma_{\rm{K^+n}}$, if one includes the charge-exchange
interaction. How this cross section is modified in the medium is
unknown.

The principal problem for extracting information on in-medium
properties of the \kp from data measured in heavy-ion collisions
is the fact that almost all observables depend simultaneously on
both, $\omega({\bf{k}}=0,\rho)$ and $\sigma_{\rm{K^+N}}^{\rm
medium}$. The situation becomes even more complicated because the
\kp are mostly produced  by $\Delta N$ collisions with an
experimentally not accessible cross section. Furthermore, the life
time of the $\Delta$ in medium is not known. Different values for
these little known quantities can lead to very similar values of
the \kp observables because the effects compensate. For example,
lowering of the in-medium \kp mass, Eq.~(\ref{kpmass}), has to a
large extend the same effect as the lowering of the in-medium
N$\Delta$ cross section. This has created quite a confusion in the
past.

Studies have been made to extract the \kp nucleus potential from
\kp spectra observed in $\pi A$  reactions
\cite{Benabderrahmane:2008qs}. Pions are absorbed close to the
surface and therefore the \kp observed in these reactions come
from regions with a moderate density where the \kp nucleus
potential is weak. In addition, the uncertainty of the involved
$\rm{B B} \to$ \kp $\rm{Y N}$ cross sections, where B stands for
baryons and Y for hyperons, and of the resonant $\pi$ N $\to$ \kp
X cross sections do not cancel exactly in the ratio of cross
sections, when plotted as a function of the momentum. Consequently,
changes in the potential can modify the results in a similar way
than changes of the cross sections. Therefore one has to wait
until this approach has been proven as robust against variation
of the little known input parameters as far as the determination
of the potential parameters is concerned.

Extensive studies of the different  \kp observables have revealed
that {\it one} \kp observable can be identified, which does not
depend simultaneously on several of these little known quantities.
This is the ratio of the momentum spectra at small transverse
momentum measured for symmetric systems of different sizes. We
show that this observable allows for the determination of the
``in-medium mass'', $\omega({\bf{k}}=0,\rho)$, if the ratio is
measured for a light and a medium mass symmetric system. It allows
as well for the determination of  the in-medium KN scattering
cross section, $\sigma_{\rm{K^+ N}}^{\rm medium}$, if the ratio is
measured for symmetric medium mass and heavy systems. Using ratios
of the spectra many of the above mentioned uncertainties cancel or
are minimized.

These results have been obtained with the IQMD program, an event
generator which simulates heavy-ion reactions from the initial
separation of projectile and target up to the final distribution
of fragments, nucleons and mesons. The details of IQMD program,
especially how strange particles are produced and propagated
 in this approach have been
extensively described in Ref.~\cite{Hartnack:1997ez,PR}. In the
standard set up, which is used for this studies if not otherwise
mentioned, we employ a momentum-dependent soft hadronic equation
of state~\cite{Hartnack:2005tr,PR}, the $\rm{N N} \to \rm{N N}
\to$ \kp cross section of Sibirtsev \cite{Sibirtsev:1995xb} and
the $\rm{NN} \to \rm N \Delta \to$ \kp cross section of Tsushima
et al.~\cite{Tsushima:1998jz,Tsushima:1994rj}.

\begin{figure}[b]
\psfig{figure=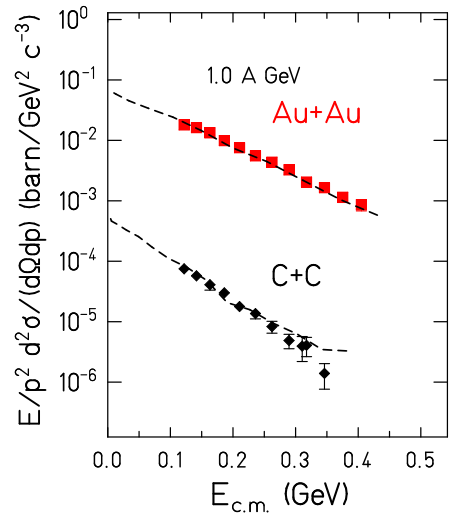,width=0.4\textwidth}
\caption{Inclusive spectra of \kp in reactions of Au+Au and C+C at
1 \AGeV incident energy taken at $\theta_{\rm{lab}}=44^\circ$ in
comparison to IQMD calculations \cite{PR}.
   } \label{Kplus}
\end{figure}

A necessary prerequisite for such studies is the successful
description of the experimentally measured \kp spectra.
Figure~\ref{Kplus} shows the spectrum of \kp mesons as a function
of the energy in the center of mass for C+C and Au+Au collisions
at 1.0 \AGeV. The experimental results \cite{Forster:2007qk} are
compared with IQMD calculation using the standard setup
($\sigma^{\rm{free}}(\rm{K^+N}\to \rm{K^+N})=\sigma^{\rm
medium}(\rm{K^+N}\to \rm{K^+N}), \alpha_{\rm K^+} = 0.08$) of
these calculations.
The absolute yield as well as the form of the spectra are well
described by our simulation program~\cite{PR}.

\begin{figure}[b]
\psfig{figure=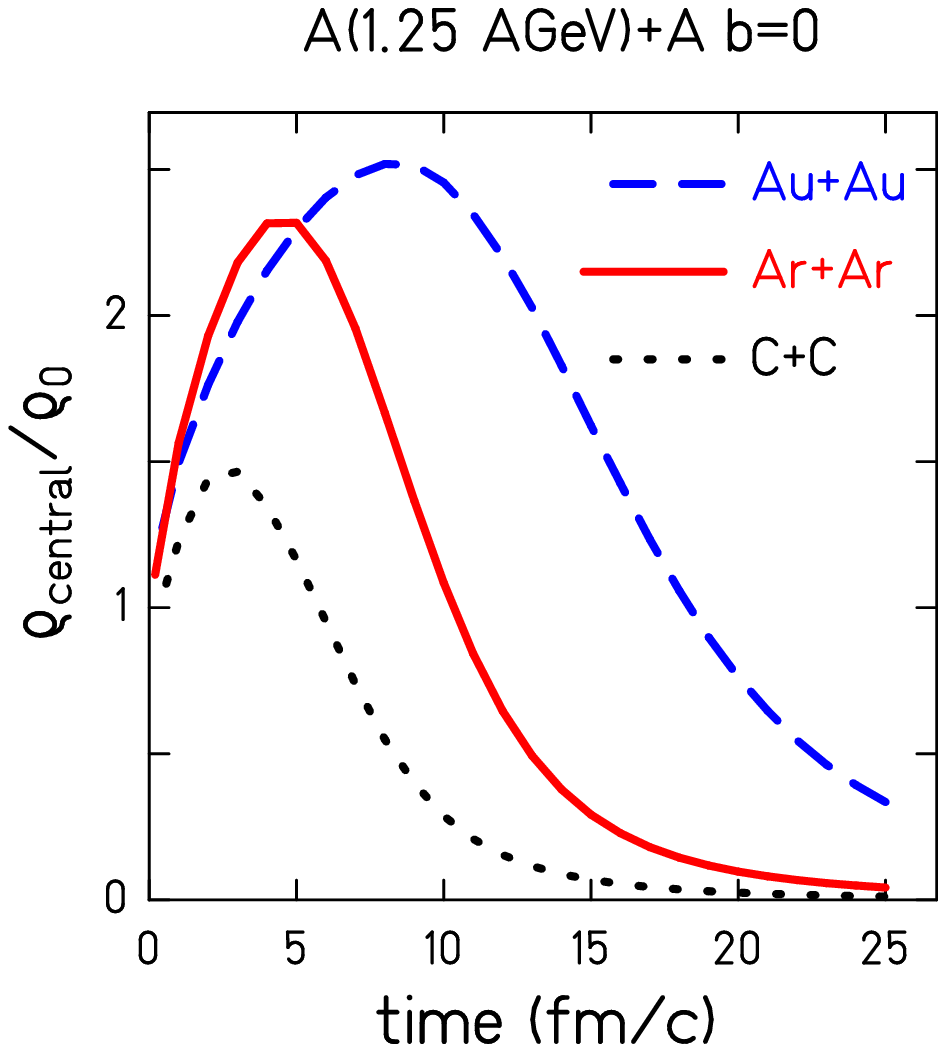,width=0.4\textwidth}
\psfig{figure=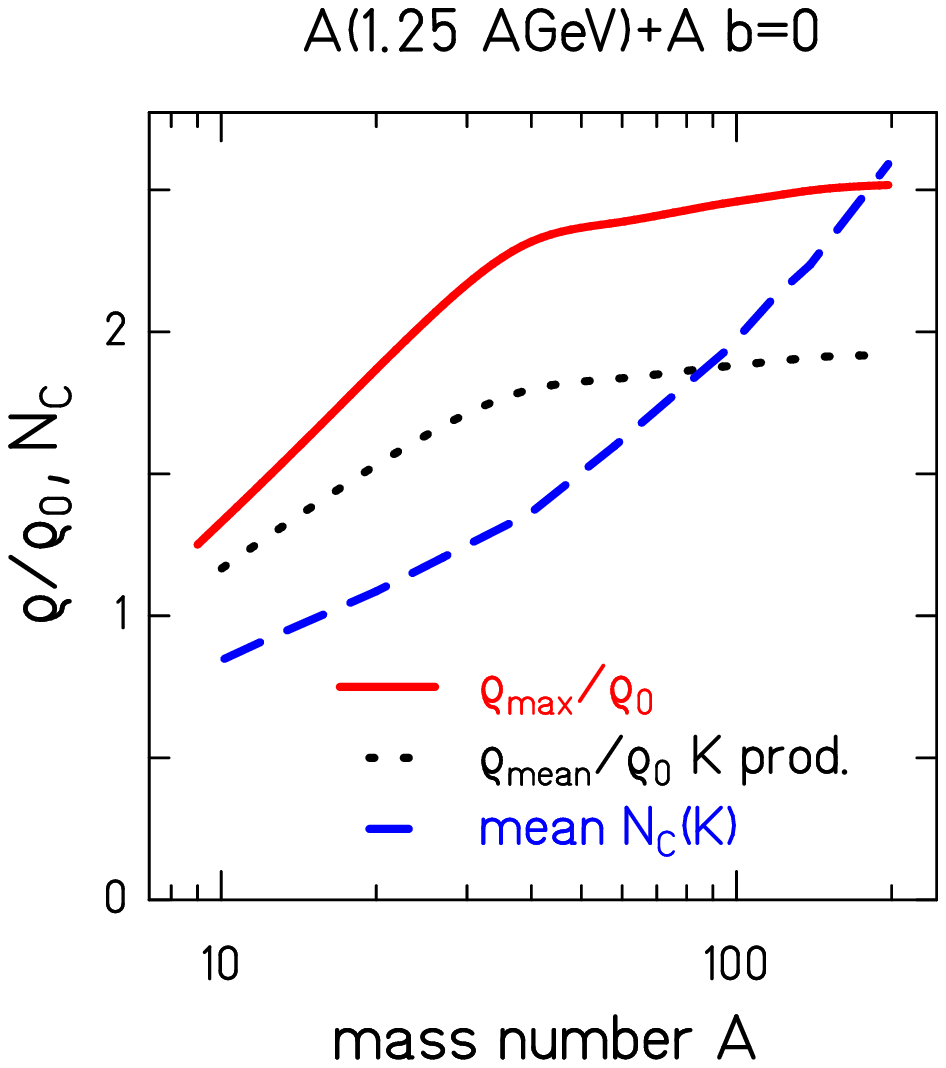,width=0.4\textwidth} \caption{Top:
Central density obtained in different symmetric $A+A$ systems at
1.25 \AGeV as a function of time. Bottom:  Maximum of the central
density and of the average density where kaons are produced, as well as of
the average number of rescattering collisions of a kaon, $N_C(\rm K)$, in the medium
 as a function of the projectile mass $A$ in symmetric systems $A+A$ at 1.25 \AGeV incident energy.
 } \label{dens}
\end{figure}

The basis for the possibility to extract both,
$\sigma^{\rm{medium}}(\rm{K^+N}\to\rm{K^+N})$ and $\alpha_{\rm
K^+}$, from the same observable is the density profile in
symmetric systems of different size, shown in Fig.~\ref{dens}.
\
The upper panel shows the time evolution of the central density of
the system obtained in symmetric reactions of Au+Au (dashed line),
Ar+Ar (full line) and C+C (dotted line) at  an incident energy of
1.25 \AGeV. The maximum density reached in the system increases
strongly when going from C+C to Ar+Ar but changes only moderately
when continuing to Au+Au. However, the duration of the
high-density phase increases strongly when comparing Ar+Ar and
Au+Au. The lower panel refines this analysis by presenting the
maximum value of the central density (full line) as a function of
the projectile mass number $A$ in symmetric reactions $A+A$ at
1.25 \AGeV incident energy. The density rises strongly up to about
$A=40$ and then only slightly for the higher masses. A similar
conclusion can be drawn when looking on the mean density at the production
points of kaons (dotted line). These densities
enter directly into the \kp nucleus potential and have an
important influence on the production yield of the kaons. Again,
the yield rises up to $A=40$ and then saturates at higher masses.
In contrast, the mean number of collisions $N_C(\rm K)$ which a
kaon suffers before leaving the system (dashed line), rises quite
moderately with $A$ up to $A=40$ and then increases much stronger
for heavier systems. Therefore, the large mass region is the realm
for measuring $ \sigma^{\rm{medium}}(\rm{K^+N}\to \rm{K^+N})$.
Below projectiles of mass 40 we see a strong increase of the
densities with system size but few rescattering collisions because
the systems are so small that rescattering does not become
important. This is the realm for measuring $\alpha_{\rm K^+}$.

\begin{figure}
\psfig{figure=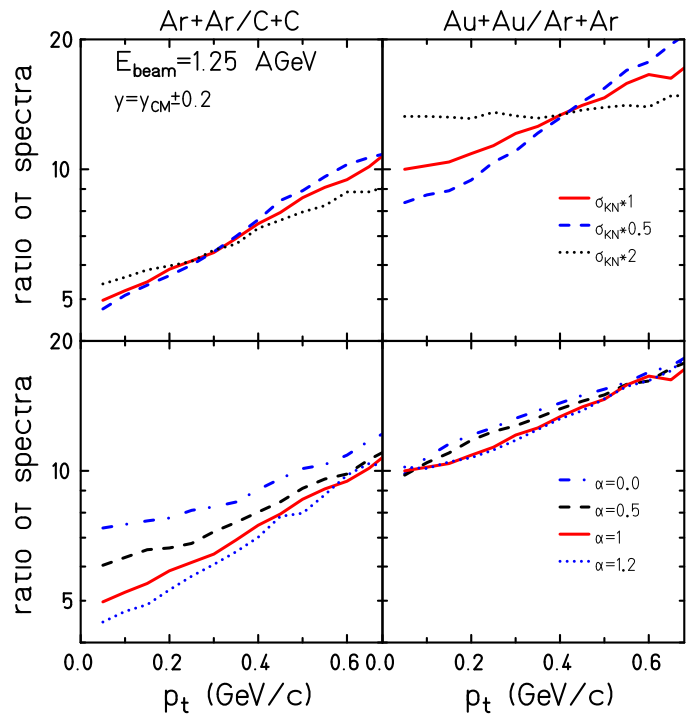,width=0.5\textwidth} \caption{Ratio of
transverse momentum spectra for symmetric systems of different
sizes as a function of the $K^+N$ elastic rescattering cross
section (top)  and of the \kp nucleus potential strength (bottom).
On the left hand side we display this ratio for C+C and Ar+Ar, on
the right hand side for Au+Au and Ar+Ar.} \label{fig2}
\end{figure}

Figure~\ref{fig2} shows the result of our calculations. It
presents the ratio of the transverse momentum spectra close to
midrapidity, obtained in Ar+Ar and C+C collisions (left) and Au+Au
and Ar+Ar collisions (right). The top panels show the influence of
the rescattering cross section. The free KN rescattering cross
section has been multiplied by a coefficient between 0.5 and 2
while leaving all the other parameter unchanged. The bottom panel
displays the variation of this ratio with the change of the
strength of the \kp nucleus potential by applying a factor
$\alpha$ to the potential constants $\Sigma_s$, $\bf{\Sigma_v}$
and $\Sigma_v^0$ of Eq.~(\ref{schaf}).
As expected from the discussion above, the ratio of the yields for
the two smaller systems is almost independent of the rescattering
cross section but depends strongly on the strength of the \kp
nucleus potential. Rescattering is not very frequent in these
light systems, therefore its  influence on the spectrum is
moderate. The \kp nucleus potential, in the contrary, has a direct
influence on the spectra, as  shown in the lower panels. The
maximal density as well as the density profile is different for
the both systems and due to this difference the \kp nucleus
potential acts differently. Whereas in C+C even the central
density does not exceed much the normal nuclear matter density, in
the Ar+Ar system the densities exceeds already twice normal
nuclear matter density. By comparing the lighter systems one cannot learn
much on the rescattering cross section but the spectra becomes
sensitive to the strength of the \kp nucleus potential. The slope
of the ratio is a direct measure of this strength and present-day
experiments are sufficiently precise to extract the strength of
the potential.

If one compares the two heavier systems (right panel of
Fig.~\ref{fig2}, a completely different scenario emerges.
Rescattering becomes very important in the Au+Au reactions where
almost all \kp undergo rescattering. Therefore, the influence of
the rescattering cross section on the spectra is very visible as
shown in the upper right panel. On the other side, the ratio is
hardly influenced by the strength of the \kp nucleus potential
because
the density at which the \kp have their last collisional
interaction with the surrounding nucleons is rather similar and
therefore, the \kp nucleus  potential acts in a very similar way
on the \mbox{K$^+$}, leaving the ratio unchanged.

\begin{figure}
\psfig{figure=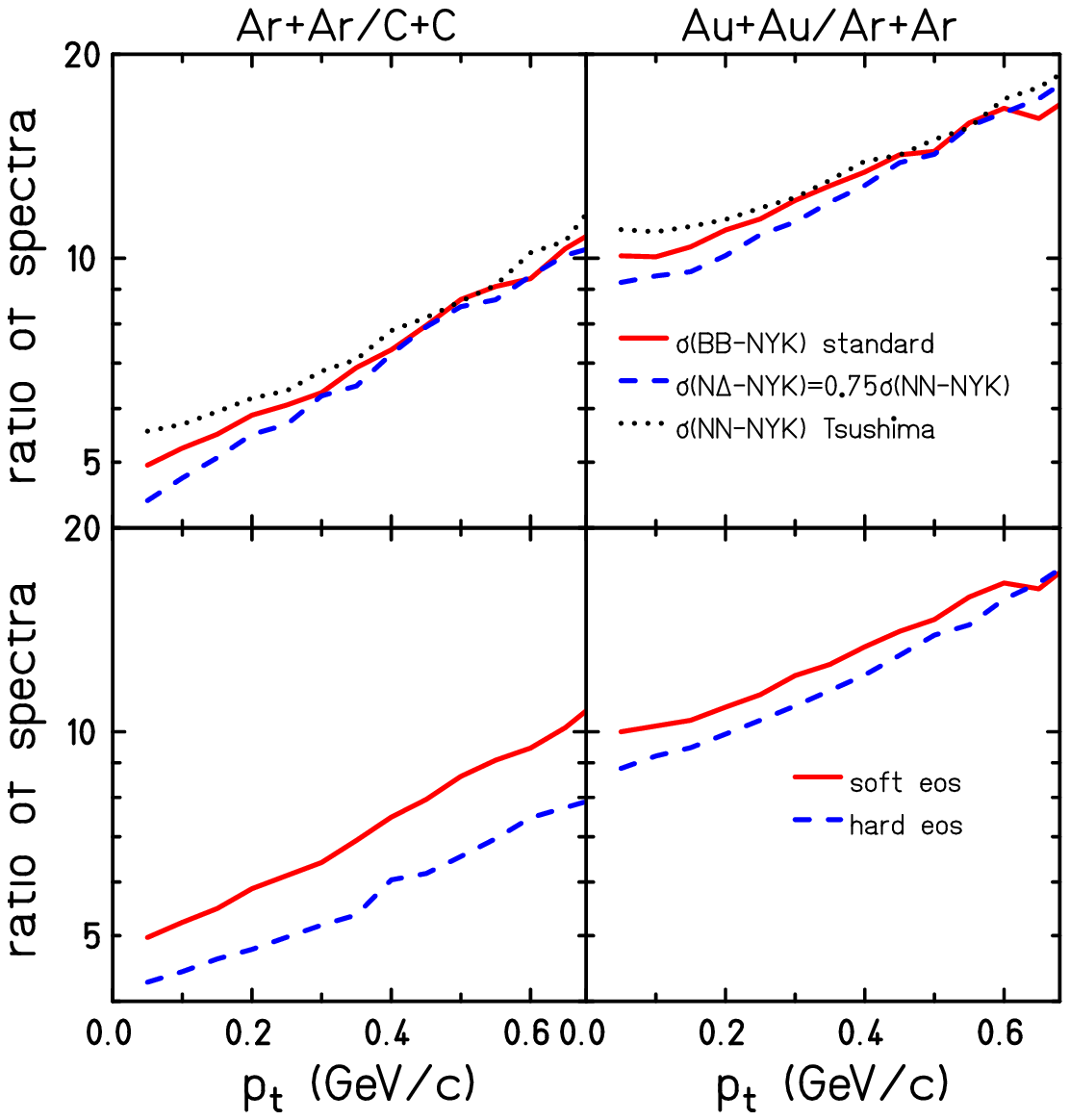,width=0.5\textwidth} \caption{Ratio of
transverse momentum spectra for symmetric systems of different
sizes as a function  of different \kp production cross  sections
(top)  and of the nuclear equation of state (bottom). On the left
hand side we display this ratio for C+C and Ar+Ar, on the right
hand side for Au+Au and Ar+Ar.} \label{fig3}
\end{figure}

Before robust conclusions can be drawn from this observation it is
necessary to verify that the other little known input quantities
do not spoil this result. In Fig.~\ref{fig3} we display how
other 
input quantities might influence the ratio of the spectra. In the
top panels the influence of the \kp production cross sections and
in the bottom panels that of the hadronic equation of state is
shown. Replacing the standard $\rm N \Delta \to \rm{N N K^+}$
cross section by an isospin corrected $\rm N \Delta \to \rm{N N
K^+}$ cross section \cite{Randrup:1980qd} or replacing the
standard $\rm{N N} \to \rm{N N K^+}$ cross section by the
parametrization of Tsushima et
al.~\cite{Tsushima:1998jz,Tsushima:1994rj}, in which the final
state interaction is not eliminated, changes only the absolute
value of the ratio of the spectra but does not harm the slope of
the ratio as a function of $p_{\rm t}$. The same is true if one
replaces the standard soft hadronic equation of state by a hard
equation of state, although a recent analysis showed that a hard
equation of state cannot be reconciled with the experimental data
\cite{Hartnack:2005tr}.
Thus other little known or unknown input quantities do not render
our conclusions useless.

Using the IQMD transport model which has shown to successfully
described the production of \kp in the range of incident energies
around 1 - 2 \AGeV, we have presented that the ratio of transverse
momentum spectra from C+C, Ar+Ar and Au+Au collisions can be used
to extract both the ``in-medium'' mass of the \kp and its
rescattering cross section. The ratio of the medium-mass collision
system and the light one is essentially sensitive to the K N
potential, while the ratio of the heavy and the medium-mass system
yield the scattering cross section. The reason is the density
reached in these collisions: It rises strongly from the light to
the medium-mass system and then only very little up to the heavy
system. The number of scatterings of the \kp, however, continues
to rise with $A$.

By comparing both ratios, it is possible to determine the \kp N
rescattering cross section as well as the \kp nucleus potential
experimentally. This opens the way to study experimentally the
properties of \kp in a nuclear environment in a direct way and
therefore, to solve one of the important outstanding questions in
hadronic matter physics.

\end{document}